\documentclass[aps,pra,nofootinbib,twocolumn,floatfix,showpacs,superscriptaddress]{revtex4-2}
\usepackage[utf8]{inputenc}
\usepackage{amsmath,graphicx,epsfig,amsthm,color,bm,braket}
\usepackage{pifont,bbding,amssymb,soul,mathrsfs}
\usepackage{indentfirst}
\usepackage{times,dsfont,units}
\usepackage{appendix, comment}

\usepackage[hyperindex,breaklinks,colorlinks=true,citecolor=blue,urlcolor=blue]{hyperref}

\usepackage{ulem}

\newcommand{\Tr}{\text{Tr}}

\usepackage{dcolumn}

\usepackage[hyperindex,breaklinks,colorlinks=true,citecolor=blue,urlcolor=blue]{hyperref}

\begin{document}
\title{Multiple-Noise-Resilient Nonadiabatic Geometric Quantum Control of Solid-State Spins in Diamond}

\author{Si-Qi Chen}
\affiliation{School of Physics, State Key Laboratory of Crystal Materials, Shandong University, Jinan 250100, China.}
\author{Qi-Tao Duan}
\affiliation{School of Physics, State Key Laboratory of Crystal Materials, Shandong University, Jinan 250100, China.}
\author{Chengxian Zhang}
\email{cxzhang@gxu.edu.cn}
\affiliation{School of Physical Science and Technology, Guangxi University, Nanning 530004, China}
\affiliation{Guangxi Key Laboratory for Relativistic Astrophysics, School of Physical Science and Technology, Guangxi University, Nanning 530004, China}
\author{He Lu}
\email{luhe@sdu.edu.cn}
\affiliation{School of Physics, State Key Laboratory of Crystal Materials, Shandong University, Jinan 250100, China.}

\begin{abstract}
Reliable and robust control lies at the core  of implementing quantum information processing with diamond nitrogen-vacancy (NV) centers. However, control pulses inevitably introduce multiple errors, leading to decoherence and hindering scalable applications.  Here, we experimentally report an experiment-friendly multiple-noise-resilient nonadiabatic geometric quantum gate~(MNR-NGQG) that can significantly improve conventional dynamical gate in both robustness and coherence. Notably, even when the detuning fluctuation range is comparable to the maximum Rabi frequency, the single-qubit gate performance of the MNR-NGQG remains almost unchanged. Besides, the coherence time of the electron spin is significantly extended to 690 $\pm$ 30 $ \mu$s,  3.5 times that of the naive dynamical counterpart. As a result, the fidelity of single-qubit gates reaches 0.9992(1), as characterized by quantum process tomography.   With its experimentally feasible design and relaxed hardware requirements, our work offers a solid paradigm for achieving high-fidelity quantum control in NV center system, paving the way for practical applications in quantum information science.

\end{abstract}
\maketitle

The nitrogen-vacancy (NV) center in diamond is a point defect consisting of a substitutional nitrogen atom adjacent to a vacancy in the diamond lattice~\cite{DOHERTY2013PRReports}. The central electron spin of NV center exhibits long coherence times at room temperature~\cite{Abobeih2018NC} and can be optically initialized and readout~\cite{Robledo2011Nature,Robledo2011NJP}. The electron spin is capable to couple the intrinsic nuclear nitrogen spin~($^{14}$N or $^{15}$N) as well as the surrounding nuclear carbon-13 spins~($^{13}$C) in the host diamond lattice, making it a promising solid-state spin register for quantum information processing~\cite{Awschalom2018NP}. The full control over NV-based spin register has been experimentally realized to create multi-spin entanglement with spins up to 10~\cite{Kolkowitz2012PRL,Taminiau2012PRL,Dam2019PRL,Bradley2019PRX,Randall2021Science}. Potential applications of small-scale multi-spin register have been demonstrated in quantum network~\cite{Bernien2013Nature,Hensen2015Nature,Pompili2021Science,Hermans2022Nature}, quantum error correction~\cite{Waldherr2014Nature,Cramer2016NC}, fault-tolerant operation~\cite{Rong2015NC,Abobeih2022nature}, quantum algorithm~\cite{Zhang2020PRL} and quantum simulation~\cite{Wang2015ACSNano,Randall2021Science}. The register size can be further scaled up via dipolar-mediated~\cite{Neumann:2010Nature}, photon-mediated~\cite{Auer2016PRB,Burkard2017PRB,Hannes2024} and magnon-mediated entangling gates~\cite{Fukami2021PRXQ,Ullah2022PRR,Jiang2024PRevApplied}, making NV center an appealing candidate for the noisy intermediate-scale quantum device~\cite{Preskill2018quantum}. 

Fast and robust quantum control of NV system is a fundamental requirement for the realization of advanced quantum technologies. Impressive progresses have been made in control and manipulation of spin qubits, enabling high-fidelity single-qubit~\cite{Rong2015NC,Vallabhapurapu2023PRA,Bartling2024PRApplied} and two-qubit gates~\cite{Xie2023PRL,Bartling2024PRApplied} that mitigate decoherence induced by the nuclear spin bath~\cite{Dobrovitski2009PRL,Zhao2011PRL,Casanova2016PRL,Chen2023PRB,Ungar2024PRXQ,Maile2024PRB}. However, control pulses would inevitably introduce errors, which can be categorized into the qubit frequency detuning error and Rabi frequency error \cite{Rong2015NC,Xu2020PRL} (hereafter, named as detuning and Rabi error for short). Geometric gates exhibit robustness against the noise because geometric phases depend solely on the overall evolution paths rather than the specific Hamiltonian of quantum systems~\cite{Pachos1999PRA,ZANARDI1999PLA,Ekert2000JMO,sjoqvist2008Physics,Sjoqvist2015IJQC}. Along this spirit, adiabatic geometric gates are designed and demonstrated~\cite{ZANARDI1999PLA,Pachos1999PRA,Pachos1999PRA,Falci2000Nature,Duan2001Science,Solinas2003PRA,Faoro2003PRL,Toyoda2013PRA,Wu2013PRA,Huang2019PRL,Jones2000Nature}, in which the quantum system evolves along adiabatic and cyclic paths, rendering it robust to errors in control parameters. However, adiabatic process is generally time-consuming, thereby reducing gate fidelity due to decoherence~\cite{Pachos1999PRA,Jones2000Nature,Duan2001Science}. Nonadiabatic geometric quantum gates~(NGQGs) have been proposed to speed up the adiabatic process~\cite{Wang2001PRL,Zhu2002PRL,Sjoqvist2012NJP,Xu2021PRL}, and have been realized in NV center~\cite{Zu2014Nature,Arroyo2014NC,Sekiguchi2017NaturePhotonics,Zhou2017PRL,Nagata2018NC,Kleisler2018Npj,Dong2021PRApplied} as well as other platforms~\cite{AbdumalikovJr2013Nature,Feng2013PRL,Song2017NC,Xu2018PRL,Ishida2018OL,Yan2019PRL,Zhang2019NJP,Xu2020PRL,Ai2020PRApplied,Li2021PRApplied,Ma2023PRApplied,Yang2023PRApplied}. Conventional NGQG focuses on the robustness against either Rabi or detuning errors~\cite{Yan2019PRL,Xu2020PRL,Ai2020PRApplied,Li2020PRResearch,Dong2021PRApplied,Li2021PRApplied,Yang2023PRApplied}, but not both simultaneously.
To suppress both types of errors simultaneously, existing NGQGs either require control pulses with multiple frequencies~\cite{Ma2023PRApplied,Liang2016PRA,Kleisler2018Npj} or are restricted to specific gate sets~\cite{Liu2020PRResearch,Zhang2021PRL}. Thus, an experiment-friendly and universal NGQG remains elusive.

In this work, we demonstrate an experiment-friendly multiple-noise-resilient nonadiabatic geometric quantum gate, i.e., MNR-NGQG~\cite{Fang2024PRA}, which refines the evolution without introducing external detuning field~\cite{Zhang2017PRL,Liu2021AQT}. Using a single-frequency driving field, this method can simultaneously suppress both Rabi and detuning errors. Notably, even when the detuning fluctuation range is comparable to the maximum Rabi frequency, the single-qubit gate performance of the MNR-NGQG remains almost unchanged. Leveraging this framework, we demonstrate high-fidelity quantum control and long coherence time in a single NV center at room temperature, which substantially outperforms the conventional dynamical gate.  Specifically,  the  reported coherence time of the electron spin has been significantly extended to 690 $\pm$ 30 $ \mu$s,  3.5 times as the naive dynamical  one. The fidelity of single-qubit gates can reach as high as 0.9992(1), as characterized by quantum process tomography.

\begin{figure}[ht!bp]
\includegraphics[width=\linewidth]{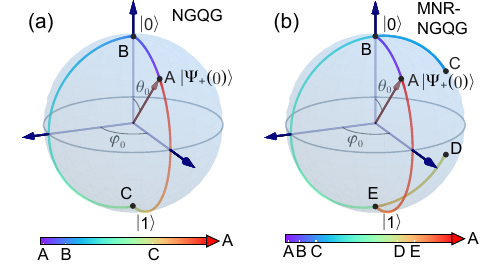}
\caption{\label{Fig:Concept}The evolution path of conventional NGQG~(a) and MNR-NGQG~(b) to realize a single-qubit rotation $R_{\bm n}(\xi)$.}
\end{figure}

For a two-level quantum system driven by a monochromatic field, the Hamiltonian~($\hbar=1$) is given by
\begin{equation}\label{Eq:Ham}
\mathcal H(t)=\frac{1}{2}\begin{pmatrix}
0&\Omega(t)e^{-i\phi(t)}\\
\Omega(t)e^{i\phi(t)}&0
\end{pmatrix},
\end{equation}
where $\Omega(t)$ denotes the time-dependent Rabi frequency, proportional to the amplitude of the driving field, and $\phi(t)$ is the phase of the driving field. The evolution of geometric gate is described in the dressed-state basis
\begin{equation}
\begin{split}
&\ket{\Psi_{+}(t)}=\cos\frac{\theta(t)}{2}\ket{0}+\sin\frac{\theta(t)}{2}e^{i\varphi(t)}\ket{1},\\
&\ket{\Psi_{-}(t)}=\sin\frac{\theta(t)}{2}e^{-i\varphi(t)}\ket{0}-\cos\frac{\theta(t)}{2}\ket{1},\\
\end{split}
\end{equation}
which can be represented by spherical coordinates~$(\theta(t), \varphi(t))$ on Bloch sphere respectively. Under the cyclic evolution condition, the dressed states acquire a global phase at final time $T$~\cite{Li2021PRResearch,Guo2023PRA}, i.e., $\ket{\Psi_{\pm}(T)}=e^{i\gamma_{\pm}}\ket{\Psi_{\pm}(0)}$ with $\gamma_{+}=-\gamma_{-}=\gamma$. Consequently, path evolution operator $U(T)$ is
\begin{equation}
    U(T)=e^{i\gamma}\ket{\Psi_{+}(0)}\bra{\Psi_{+}(0)}+e^{-i\gamma}\ket{\Psi_{-}(0)}\bra{\Psi_{-}(0)}.
\end{equation}

The global phase accumulated along a closed evolution path can be divided into two parts, i.e., the dynamical phase $\gamma_{d}$ and the geometric phase $\gamma_{g}$. The goal of NGQG is to cancel the dynamical phase $\gamma_{d}$ at the final time $T$ by appropriately designing the evolution path. In conventional NGQG, the dressed state evolves along a specific trajectory---referred as orange-slice-shaped loop---to construct a universal set of single-qubit gates. As shown in Fig.~\ref{Fig:Concept}~(a), the initial dressed state $\ket{\Psi_+(0)}$ is represented by a point $\text{A}$ with coordinate~$(\theta_0,\varphi_0)$. Cyclic evolution occurs along the path  $\text{A}\to\text{B}\to\text{C}\to\text{A}$, and the control parameters $\Omega(t)$ and $\phi(t)$ can be derived accordingly. 

\begin{figure*}[ht!bp]
\includegraphics[width=\linewidth]{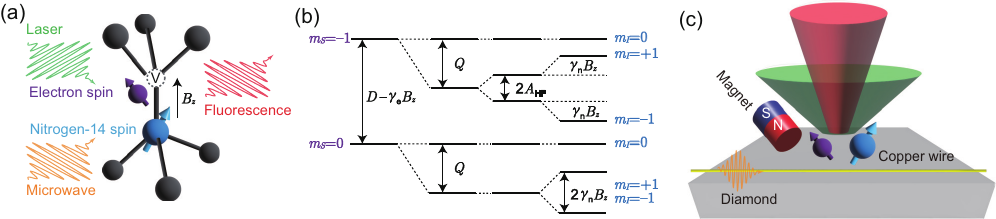}
\caption{\label{Fig:setup}(a), The structure of single N-$V$ center. (b), The configuration of energy levels of single N-$V$ center. (c), The illustration of home-built confocal microscope to initialize, manipulate and readout of spin states. The 532-nm excitation light from a diode laser is modulated via an acousto-optic modulator and then reflected by a dichroic mirror into an oil-immersion objective lens (NA = 1.25). Photoluminescence from the NV center is collected by the same objective, transmitted through the dichroic mirror, and detected by a single-photon detector. The MW field used to drive spin states is delivered via a ~20-$\mu$m-diameter copper wire connected to a printed circuit board.  }
\end{figure*} 

Next, we consider the Hamiltonian in the presence of  errors
\begin{equation}\label{Eq:Ham_error}
\mathcal H^{\prime}(t)=\frac{1}{2}\begin{pmatrix}
\delta \Omega_{\text{max}}&(1+\epsilon)\Omega(t)e^{-i\phi(t)}\\
(1+\epsilon)\Omega(t)e^{i\phi(t)}& -\delta \Omega_{\text{max}}
\end{pmatrix},
\end{equation}
where $\delta$ and $\epsilon$ represent the detuning and Rabi error, respectively—two important error sources in NV center experiments~\cite{Rong2015NC}. $\Omega_{\text{max}}$ corresponds to the maximum amplitude of driving field. Our objective is to  mitigate these two types of errors simultaneously, and thus improve the gate fidelity. To this end, the whole evolution is divided into five segments, with $\Omega(t)$ and $\phi(t)$ in each segment satisfying
\begin{subequations}\label{Eq:composite}
\begin{align}\label{Eq:composite_geo:1}
 &\int_{t_1}^{t_{2}}\Omega(t)dt=\theta_{0},\;\phi(t)=\varphi_{0}-\frac{\pi}{2}, \;t\in [t_{1}=0,t_{2}],\\
 &\int_{t_{2}}^{t_{3}}\Omega(t)dt=\frac{\pi}{3},\;\phi(t)=\varphi_{0}+\gamma +\frac{3\pi}{2},\;t\in [t_{2},t_{3}],\\
   &\int_{t_{3}}^{t_{4}}\Omega(t)dt=\frac{5 \pi}{3},\;\phi(t)=\varphi_{0}+\gamma +\frac{\pi}{2},\;t\in [t_{3},t_{4}],\\
 &\int_{t_{4}}^{t_{5}}\Omega(t)dt=\frac{\pi}{3}, \;\phi(t)=\varphi_{0}+\gamma +\frac{3\pi}{2}, \;t\in [t_{4},t_{5}],\\\label{Eq:composite_geo:5}
 &\int_{t_{5}}^{T}\Omega(t)dt=\pi-\theta_{0},~~\phi(t)=\varphi_{0}-\frac{\pi}{2}, \;t\in [t_{5},T].
\end{align}
\end{subequations}
The evolution trajectory of the MNR-NGQG is illustrated in Fig.~\ref{Fig:Concept}~(b), where the cyclic evolution occurs along $\text{A} \to \text{B} \to \text{C} \to  \text{D} \to \text{E} \to \text{A}$. The corresponding evolution operator is given by 
\begin{equation}\label{U_conventional}
\begin{split}
       U_g(T)&= U_\text{EA}(T,t_{5}) \, U_\text{DE}(t_{5},t_{4}) \, U_\text{CD}(t_{4},t_{3}) \\
       &\quad\times U_\text{BC}(t_{3},t_{2}) \, U_\text{AB}(t_{2},t_{1}) \\
       &=\cos \gamma \mathds 1_2 +i\sin \gamma 
\begin{pmatrix}
 \cos\theta_{0} & \sin\theta_{0}e^{-i\varphi_{0} }\\\sin\theta_{0}e^{i\varphi_{0} } &-\cos\theta_{0}
\end{pmatrix} \\
&=e^{i\gamma\bm{n} \cdot \bm{\sigma}},
\end{split}
\end{equation}where $\bm{n} \cdot \bm{\sigma}$ is the linear combination of Pauli matrices. Then, arbitrary single-qubit rotation $R_{\bm{n}}(\xi)$ can be realized by setting 
\begin{align}
\bm{n}=(\sin\theta_{0}\cos\varphi_{0},\sin\theta_{0}\sin\varphi_{0},\cos\theta_{0}),
\xi=-\frac{\gamma}{2}.
\end{align} 
Owing to the relaxed constraints on the amplitude function in Eq.~(\ref{Eq:composite}), we can employ pulses with smooth edges, thereby reducing the frequency bandwidth requirements for experimental devices. Besides, the external detuning field is not required, which is needed in the recent work~\cite{Liang2016PRA,Kleisler2018Npj} to suppress multiple errors.

The experiment is demonstrated with a single NV center in diamond, as shown in Fig.~\ref{Fig:setup}~(a). The NV center consists of a substitutional nitrogen atom~($^{14}$N) and a neighboring vacant site~($V$) in the diamond lattice. An additional electron is obtained from a donor within the diamond, forming a negatively charged NV center. The electron and $^{14}$N spins form a two-qubit system~(e-$^{14}$N)~\cite{van2012nature,Rong2015NC,Bradley2019PRX,Wu2019npjQI,Vallabhapurapu2023PRA,Xie2023PRL,Bartling2024PRApplied}, described by the Hamiltonian
\begin{equation}\label{Eq:Hsys}
\mathcal H_\text{NV}/2\pi=\mathcal H_\text{e}+\mathcal H_\text{n}+\mathcal H_{i},
\end{equation}
where the electronic, nuclear, and interaction components are given by
\begin{align}
\label{Eq:e}
&\mathcal H_\text{e}=DS_{z}^{2}+\gamma_{e}B_0S_{z},\\
\label{Eq:n}
&\mathcal H_\text{n}=QI_{z}^{2}+\gamma_{n}B_0I_{z},\\
\label{Eq:i}
&\mathcal H_{i}=A_\text{HF}S_{z}I_{z},
\end{align}
respectively. Here, $S_z$ and $I_z$ are the spin operators for the electron spin~($S=1$) and nuclear spin~($I=1$). Eqs.~(\ref{Eq:e}) and~(\ref{Eq:n}) describe the free
evolution of the electron and nuclear spins, respectively. The configuration of energy levels is shown in Fig.~\ref{Fig:setup}~(b). $D=2.87$~GHz is the electronic zero-field splitting~($B_z=0$) between $m_S=0$ and $m_S=\pm1$ states, and $Q=-4.95$~MHz is the nitrogen nuclear quadrupolar splitting. For $B_z\neq0$, the degeneracy of electronic~(nuclear) levels $m_S=\pm1$~($m_I=\pm1$) is lifted by $2\gamma_\text{e}B_z$~($2\gamma_\text{n}B_z$), where $\gamma_\text{e}=28$~GHz/T~($\gamma_\text{n}=-3.1$~MHz/T) is the electronic~(nuclear) gyromagnetic ratio. The hyperfine interaction
between the electron and nuclear spin is characterized by the hyperfine interaction strength $A_\text{HF}=-2.16$~MHz, as described in Eq.~(\ref{Eq:i}). The polarization and readout of spin states are achieved using a home-built confocal microscope~(Fig.~\ref{Fig:setup}~(c)), with the driving microwave~(MW) field delivered via a copper wire of approximately 20~$\mu$m diameter~\cite{Chen2025PRApplied}.

Single-qubit gates are implemented on electron spin states, where the qubit is encoded in the sublevels of $\ket{m_S=0}=\ket{0}_\text{e}$ and $\ket{m_S=-1}=\ket{1}_\text{e}$. A static magnetic field $B_z\approx 500$~G is applied along the NV symmetry axis to lift the degeneracy between the  $\ket{m_s=+1}$ and $\ket{m_s=-1}$ states by 2.827~GHz. The electronic Hamiltonian $\mathcal H_\text{e}$ interacts with an $x$-axis-polarized MW with time-dependent driving strength, i.e., $\mathcal H_{D}=\sqrt{2}\Omega(t)\cos(\omega t+\phi(t))S_x$ with $\omega=D-\gamma_\text{e}B_z-A_{\text{HF}}$. The Rabi frequency $\Omega(t)$ takes the form of 
\begin{equation}\label{Eq: MWfunction}
    \Omega(t)=\Omega _\text{max}\left [ 1-\left | \cos \left(\pi \frac{t-t_i}  {\tau_i} \right ) \right |^{n}  \right ],
\end{equation}
where $\Omega_\text{max}$ is the maximum Rabi frequency driven by the MW field and $n$ is a constant that suppresses detuning errors~\cite{Huang2019PRL}. $\tau_i$ ($t_i$) denotes the 
duration (start time) of the $i$-th segment in Eq.~(\ref{Eq:composite}). In our experiment, we set $\Omega_\text{max}=2\pi \times 4$~MHz and $n=25$.

\begin{figure*}[ht!bp]
\includegraphics[width=\linewidth]{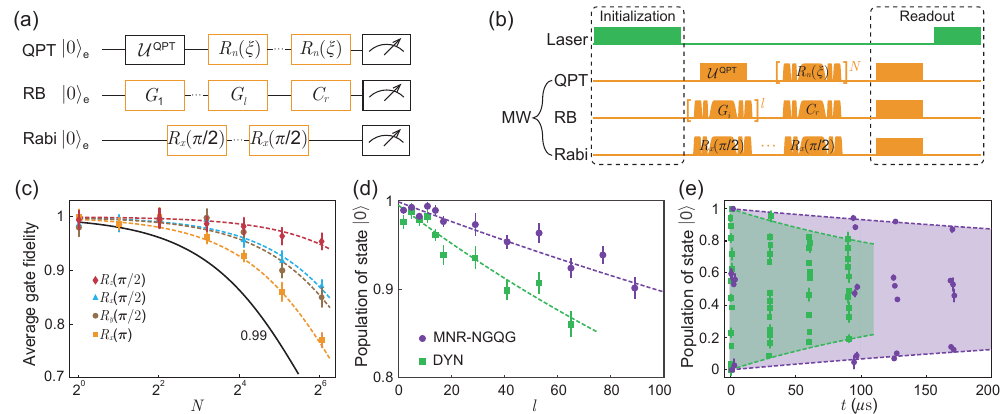}
\caption{\label{Fig:single}The quantum circuit~(a) and sequences~(b) to implement quantum process tomography~(QPT), randomized benchmarking~(RB) and Rabi oscillation.  (c), The results of average gate fidelity measured from QPT. (d), The results of fidelity estimated from RB. (e), The Rabi oscillation with MNR-NGQG.   }
\end{figure*}

We implement four single-qubit rotations using this pulse shape, i.e., $R_x(\frac{\pi}{2})$, $R_x(\pi)$, $R_y(\frac{\pi}{2})$ and $R_z(\frac{\pi}{2})$, with corresponding parameters $\left \{\theta_{0},\varphi_{0},\gamma\right \}$ given by $\left \{ \pi/2,0,-\pi/4 \right \}$, $\left \{ \pi/2,0,-\pi/2 \right \}$, $\left\{\pi/2,\pi/2,-\pi/4 \right \}$ and $\left \{ 0,\pi/2,-\pi/4 \right \}$, respectively. The MW function in each segment is determined using Eq.~(\ref{Eq:composite}). 
The performance of the single-qubit rotation is characterized via quantum process tomography~(QPT)~\cite{Brien2004PRL}. Specifically, we perform QPT to reconstruct the process matrix $\chi_{\text{exp}}$ after repeatably applying each rotation gate for $N=$1, 2, 4, 9, 17, 33 and 65 times. We then calculate the process fidelity $F_N=\Tr(\chi_{\text{exp}}\chi_{\text{ideal}})$ with respect to the ideal matrix $\chi_{\text{ideal}}$, and the results are shown in Fig.~\ref{Fig:single}~(c). The process fidelity can also be expressed as $F_{N}=1/2+1/2(1-2p)^{N}$, where $p$ is the average error per rotation gate~\cite{Dong2021PRApplied}. From this, we estimate the average gate fidelities of $R_x(\frac{\pi}{2})$, $R_x(\pi)$, $R_y(\frac{\pi}{2})$ and $R_z(\frac{\pi}{2})$ to be 0.9975(4), 0.9953(2), 0.9972(2) and 0.9992(1) respectively. The fidelity of the MNR-NGQG exceeds that of other schemes implemented in NV center~\cite{Zu2014Nature,Kleisler2018Npj,Dong2021PRApplied}. 

The MNR-NGQG is also characterized using randomized benchmarking~(RB) method~\cite{Knill2008PRA}. We initialize the electron state to $\ket{0}_\text{e}$, apply a sequences of $l$ gates $\mathcal G=\{G_1,\cdots,G_l\}$ with $G_i$ being randomly sampled from Pauli and Clifford group, and finally apply a gate $C_r$ that would return the state to $\ket{0}_\text{e}$ in the absence of noise. With noise, the probability of measuring $\ket{0}_\text{e}$ is $p_{\ket{0}_\text{e}}=[1+(1-d_{if})(1-d)^l]/2$, where $d$ is the average depolarization probability of $G_i$ and $d_{if}$ accounts for depolarization from the state preparation and measurement.  Fitting the decay $p_{\ket{0}_\text{e}}$ yields $d$, from which the average gate fidelity is calculated as $\bar{F}=1-d/2$. In our experiment, $l=\left \{ 2,5,8,11,14,17,29,41,53,65,77,89\right \}$, with 25 random sequences $\mathcal G$ generated for each $l$. The results of $p_{\ket{0}_\text{e}}$ are shown with purple dots in Fig.~\ref{Fig:single}~(d), giving an average gate fidelity $\bar{F}_\text{NGQG}=0.9989(2)$. For comparison, we implemented single-qubit rotations using a conventional naive dynamical~(DYN) approach~\cite{Kennedy2002physica,Kleisler2018Npj}. $\Omega_\text{max}$ is the same as that in MNR-NGQG. The results of $p_{\ket{0}_\text{e}}$ with DYN are shown with green squares in Fig.~\ref{Fig:single}~(d) and the average gate fidelity is $\bar{F}_\text{DYN}=0.9977(6)$, confirming the MNR-NGQG is more robust to the dynamical counterpart. The fidelity of MNR-NGQG is expected to be further improved when an isotopically purified sample is used.

The coherence protection of this scheme is verified via Rabi oscillations, where we successively apply $R_x(\pi/2)$ to $\ket{0}_\text{e}$ and measure the probability $p_{\ket{0}_\text{e}}$. As shown in Fig.~\ref{Fig:single}~(e), the DYN approach yields a coherence time of $189\pm8~\mu$s~(green squares), while the MNR-NGQG extends this to $690\pm30~\mu$s~(purple dots), approximately 3.5 times as the dynamical one.
The coherence time in the MNR-NGQG scheme also exceeds that obtained with the spin echo sequence, which yields $390 \pm 15~ \mu$s. The dashed lines in the experimental data are the envelope functions $0.5\pm0.5~\text{exp}(-t/T_{\text{coherence}})$.

\begin{figure}[htbp]
\includegraphics[width=\linewidth]{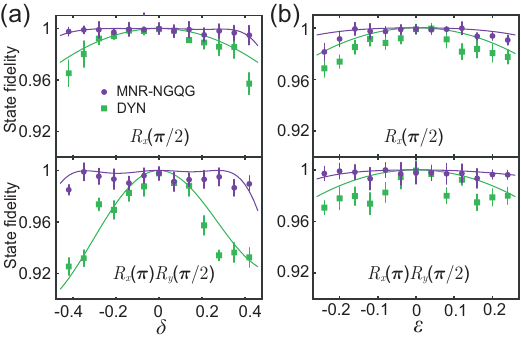}
\caption{\label{Fig:robust}The results of output state fidelity by applying $R_x(\pi/2)$, $R_x(\pi)R_y(\pi/2)$~($H$ gate), in the presence of engineered detuning error $\delta$~(a) and Rabi  error $\epsilon$~(b).}
\end{figure}

The robustness of MNR-NGQG is further validated by inducing the detuning error $\delta$ and Rabi error $\epsilon$. Experimentally, the $\delta$ is introduced by setting the  MW field frequency to $\omega^{\prime} = \omega + \delta\Omega_{\text{max}}$, while $\epsilon$ is introduced by setting MW field amplitude to yield a maximum Rabi frequency of $\Omega_\text{max}^\prime=(1+\epsilon)\Omega_\text{max}$. We implement two MNR-NGQGs---specifically, $R_x(\pi/2)$ and Hadamard gate $H=R_x(\pi)R_y(\pi/2)$---on initial states $\ket{0}_\text{e}$ respectively, and then measure the fidelity of the output state with respect to the ideal target state. As shown in Fig.~\ref{Fig:robust}, the state fidelity of the DYN gate decays more rapidly than that of the MNR-NGQG with increasing $\delta$ or $\epsilon$, MNR-NGQG can simultaneously suppress both Rabi and detuning errors. Notably, even under a detuning fluctuation range comparable to the maximum Rabi frequency, the single-qubit gate performance can exhibit negligible variation.

The design of the MNR-NGQG can be readily extended to implement two-qubit gates via frequency-selective resonant modulations. As shown in Fig.~\ref{Fig:setup}~(b), we consider a two-qubit state encoded in sublevels of $\ket{m_S=0, m_I=+1}=\ket{0}_\text{e}\ket{0}_\text{n}$, $\ket{m_S=0, m_I=-1}=\ket{0}_\text{e}\ket{1}_\text{n}$, $\ket{m_S=-1, m_I=+1}=\ket{1}_\text{e}\ket{0}_\text{n}$ and $\ket{m_S=-1, m_I=-1}=\ket{1}_\text{e}\ket{1}_\text{n}$. A two-qubit controlled-rotation~(C-ROT) gate is realized using a MW pulse~(Eq.~(\ref{Eq:composite})) with frequency of $\omega=D-\gamma_\text{e}B_z-A_{\text{HF}}$, which selectively flips the resonant spin states $\ket{0}_\text{e}\ket{0}_\text{n}\leftrightarrow\ket{1}_\text{e}\ket{0}_\text{n}$ while leaving other states unchanged. Experimentally, we implement controlled-$R_x(\pi)$ gate, i.e., $\text{C}_\text{n}$-$\text{ROT}_\text{e}=\ket{0}_\text{n}\bra{0}\otimes R_x(\pi)+\ket{1}_\text{n}\bra{1}\otimes\mathds 1_2$, where the nuclear spin acts as the control qubit and electron spin as the target qubit. We set $\Omega_{\text{max}}=2\pi \times0.7$~MHz for the realization of selective spin-state flipping, and the fidelity of $\text{C}_\text{n}$-$\text{ROT}_\text{e}$ gate is $0.950(2)$~(Details of the two-qubit gates are provided in the Appendix). Since the Rabi frequency $\Omega_{\text{max}}$ required for implementing two-qubit gate is significantly weaker than that for single-qubit gates, the noise induced by the spin bath becomes non-negligible, thereby reducing the fidelity of two-qubit gate. The fidelity of two-qubit gate is expected to be higher when implemented  between an electron spin and a $^{13}$C nuclear spin, as the $\Omega_{\text{max}}$ is significantly larger in such system~\cite{Zu2014Nature,Huang2019PRL}.  

\begin{figure*}[htbp]
\includegraphics[width=\linewidth]{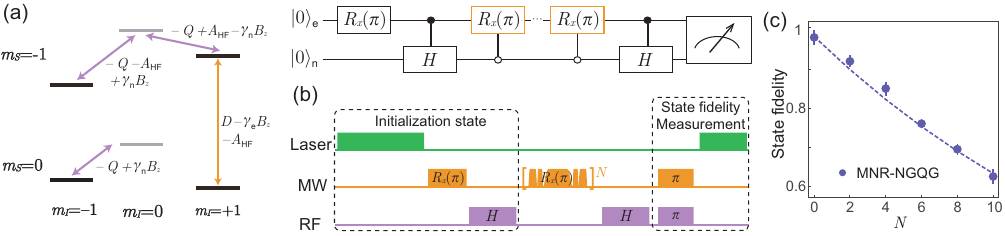}
\caption{\label{Fig:two_qubit}(a) The energy levels used to implement the two-qubit gate. (b) The quantum circuit and pulse sequences used to evaluate the performance of the two-qubit gate. (c) Experimental results of the output-state fidelity as a function of $N$.}
\end{figure*}

In conclusion, we experimentally demonstrate an experiment-friendly MNR-NGQG with solid-state spins in a single NV center, which is implemented without introducing external detuning field. This scheme enables simultaneous suppression of both detuning and Rabi errors arising from control imperfections, with the single-qubit gate performance of the MNR-NGQG remaining nearly unaffected even when the detuning fluctuation range is comparable to the maximum Rabi frequency. In addition, the coherence time of the electron spin is greatly enhanced from 189 $\pm$ 8 $ \mu$s to 690 $\pm$ 30 $ \mu$s. Thus,  the fidelity of single-qubit gate can reach as high as 0.9992(1), as characterized by quantum process tomography. 	On the other hand, the average single-qubit gate fidelity via randomized benchmarking is 99.89(2)\%, exceeding the threshold required for implementing state-of-the-art error correction codes such as surface codes~\cite{kitaev_quantum1997,Fowler2012PRA}. 
Higher fidelity is expected, especially when implementing the scheme experimentally on an NV center in an isotopically purified sample, and with a large $\Omega_{\text{max}}$ that can suppress detuning errors. Our work provides an experimentally feasible design for implementing quantum gates in NV centers with relaxed hardware requirements, facilitating high-fidelity multi-qubit operations. This advancement holds significant promise for applications in quantum information processing based on solid-state spin systems.

\appendix
\section{The realization of two-qubit gate with MNR-NGQG} \label{A1:two-qubit}
The energy levels used to implement the two-qubit gate are marked with black bold lines in Fig.~\ref{Fig:two_qubit}~(a). The two-qubit gate is realized using a resonant MW-driven field with frequency $\omega = D - \gamma_e B_z -A_{\text{HF}}$, which matches the energy gap between $\ket{0}_{\text{e}} \otimes \ket{0}_{\text{n}}$ and $\ket{1}_{\text{e}} \otimes \ket{0}_{\text{n}}$. This field selectively manipulates the electron spin conditioned on the nuclear spin state, i.e., it induces a rotation of electron spin when the nuclear spin state is in $\ket{0}_{\text{n}}$, while leaving the electron spin unchanged when the nuclear spin is in $\ket{1}_{\text{n}}$. Accordingly, the control Hamiltonian is written as
\begin{equation}
\mathcal{H}^{\text{two}}_{D} = \sqrt{2} \, \Omega(t) \cos\bigl(\omega t + \phi(t)\bigr) 
 \sigma_x^{(\text{e})} \otimes \ket{0}_{\text{n}}\bra{0}_{\text{n}} .
\end{equation}
$\text{C}_\text{n}$-$\text{ROT}_\text{e}=\ket{0}_\text{n}\bra{0}\otimes R_x(\pi)+\ket{1}_\text{n}\bra{1}\otimes\mathds 1_2$ is realized by setting parameters $\left\{\theta_{0}, \varphi_{0}, \gamma\right\}$~(Eq.~(\ref{Eq:composite})) to be $\left\{ \pi/2, 0, -\pi/2 \right\}$. 

To estimate the fidelity of the two-qubit gate, we repeatably apply $\text{C}_\text{n}$-$\text{ROT}_\text{e}$ $N$ times to initial state $1/\sqrt{2}(\ket{0}_{\text{n}}+\ket{1}_{\text{n}}) \otimes \ket{1}_{\text{e}}$, and calculate the gate fidelity according to the decay of the output-state fidelity with increasing $N$~\cite{Chow2012PRL}. As shown in Fig.~\ref{Fig:two_qubit}~(b), the initial state $1/\sqrt{2}(\ket{0}_{\text{n}}+\ket{1}_{\text{n}}) \otimes \ket{1}_{\text{e}}$ is prepared by applying an $R_x(\pi)$ rotation to the electron spin and a Hadamard gate to the nuclear spin. Then, the two-qubit gate is repeatedly applied to this initial state $N$ times~(where $N$ is an even number), followed by a second Hadamard gate. Finally, we measure the fidelity of output state with respect to $\ket{1}_{\text{e}}\otimes \ket{0}_{\text{n}}$. The Hadamard operation on the nuclear spin is implemented via a holonomic quantum gate driven by two resonant RF pulses, with frequencies $-Q-A_{\text{HF}}+\gamma_nB_z$ and $-Q+A_{\text{HF}}-\gamma_nB_z$~\cite{Arroyo2014NC}, respectively. As shown in Fig.~\ref{Fig:two_qubit}~(c), the gate fidelity is calculated by fitting the data to function $F_N = A F_{g}^{N} + B$, where $A(B)$ is fitting parameter, and $F_g$ denotes the gate fidelity.

\bibliography{NGQG}

\end{document}